\documentclass[11pt]{article}
\usepackage[T1]{fontenc}
\usepackage[latin1]{inputenc}
\usepackage{avant}
\usepackage{geometry}
\usepackage{graphics}
\usepackage{setspace}
\makeatletter

\newcommand{\LyX}{L\kern-.1667em\lower.25em\hbox{Y}\kern-.125emX\spacefactor1000}

\makeatother

\begin{document}

\title{Ergodicity and the Classical \protect\( \lambda \phi ^{4}\protect \) Lattice
Field Theory\thanks{
Study supported by the undergraduate PIBIC/CNPq research program, Brazil (to
be presented at the \emph{XIX Encontro Universitário de Iniciação à Pesquisa},
UFC)
}}

\author{Artur B. Adib\thanks{
email: adib@fisica.ufc.br
}\\
Advisor: Carlos A. S. de Almeida\thanks{
email: carlos@fisica.ufc.br
}}

\date{\emph{Departamento de Física, Universidade Federal do Ceará, Brazil}}

\maketitle
\begin{abstract}
In this talk we present some studies in the approach to equilibrium of the classical
\( \lambda \phi ^{4} \) theory on the lattice, giving particular emphasis to
its pedagogical usefulness in the context of classical statistical field theory
(such as both the analytical and numerical evaluation of correlation functions),
as well as in the context of statistical mechanics (such as the equivalence
of ensembles in the thermodynamic limit). Since the quartic term could be regarded
as a smooth perturbation to the (integrable) gaussian theory, we also discuss
the connection of our results with the KAM theorem, showing that the ergodic
and non-ergodic regimes are not sharply separated.
\end{abstract}

\section{Introduction}

There has been some interest recently in classical fields as effective theories
for finite temperature quantum field theory (see, for instance, \cite{gleiser, aarts}).
Furthermore, phase transitions, thermalization, ergodicity and all that from
statistical mechanics are still topics of much research in the context of field
theories. This talk will be based mostly on Aarts' paper \cite{aarts2}, which
presents some results in the thermalization of the \( \lambda \phi ^{4} \)
theory on the lattice for various ensembles. We will be interested, in particular,
in deriving analytical results related to the canonical ensemble in equilibrium
and comparing them with the results of lattice simulations out-of-equilibrium,
showing how the thermalization of the theory depends on \( \lambda  \). This
latter result is discussed in connection with the KAM theorem \cite{parisi, casetti},
which in the context of the Fermi-Pasta-Ulam problem has already been studied
by Parisi elsewhere \cite{parisi2}. Although the simulations are done in a
closed microcanonical system, in the thermodynamic limit we recover the equivalence
of the ensembles \cite{parisi}, therefore justifying the derivation of the
correlation functions with the canonical probability measure.

\section{The Classical \protect\( \lambda \phi ^{4}\protect \) Theory}

We will adopt the traditional \( \lambda \phi ^{4} \) action in 1+1 dimensions:
\begin{equation}
\label{action}
S=\int dt\int _{0}^{L}dx\left[ \frac{1}{2}\left( \partial _{t}\phi \right) ^{2}-\frac{1}{2}\left( \partial _{x}\phi \right) ^{2}-\frac{1}{2}m^{2}\phi ^{2}-\frac{1}{8}\lambda \phi ^{4}\right] .
\end{equation}
 The equation of motion for this action could then be obtained directly using
the field version of the Euler-Lagrange equation:
\begin{equation}
\label{euler_lag}
\frac{\partial {\cal L}}{\partial \phi }-\frac{\partial }{\partial x}\frac{\partial {\cal L}}{\partial (\partial _{x}\phi )}-\frac{\partial }{\partial t}\frac{\partial {\cal L}}{\partial (\partial _{t}\phi )}=0,
\end{equation}
where \( {\cal L} \) is the lagrangian density:
\begin{equation}
\label{lag_density}
{\cal L}=\frac{1}{2}\left( \partial _{t}\phi \right) ^{2}-\frac{1}{2}\left( \partial _{x}\phi \right) ^{2}-\frac{1}{2}m^{2}\phi ^{2}-\frac{1}{8}\lambda \phi ^{4}.
\end{equation}
Then the equation of motion is:
\begin{equation}
\label{eq_motion}
\partial ^{2}_{t}\phi =\partial _{x}^{2}\phi -m^{2}\phi -\frac{1}{2}\lambda \phi ^{3}.
\end{equation}
The energy density is also easily obtained as the Legendre transform of \( {\cal L} \):
\[
{\cal E}=\pi \partial _{t}\phi -{\cal L}=\frac{1}{2}\pi ^{2}+\frac{1}{2}\left( \partial _{x}\phi \right) ^{2}+\frac{1}{2}m^{2}\phi ^{2}+\frac{1}{8}\lambda \phi ^{4},\]
whence we obtain the energy functional:
\begin{equation}
\label{energy}
E[\phi ,\pi ]=\int _{0}^{L}dx\left[ \frac{1}{2}\pi ^{2}+\frac{1}{2}\left( \partial _{x}\phi \right) ^{2}+\frac{1}{2}m^{2}\phi ^{2}+\frac{1}{8}\lambda \phi ^{4}\right] ,
\end{equation}
with \( \pi =\partial {\cal L}/\partial \dot{\phi }=\partial _{t}\phi  \) the
conjugate momentum to \( \phi  \).

\section{Equilibrium Properties}

We start our statistical study by writing down the equilibrium partition function
for the classical scalar field theory:
\begin{equation}
\label{part_functional}
{\cal Z}=\int d\pi \int d\phi \exp \left( -\beta E[\phi ,\pi ]\right) ,
\end{equation}
where the functional measures \( d\pi  \) and \( d\phi  \) are the continuum
limiting cases of, respectively,
\[
d\pi =\prod _{i}d\pi _{i},\: \: \: \: d\phi =\prod _{i}d\phi _{i}\]
Since the \( \lambda \phi ^{4} \) term makes our \( \phi  \) integral non-gaussian,
we shall analyze expectation values for the (gaussian) canonical momentum \( \pi  \):
\[
\left\langle \pi (x)\right\rangle _{T}=\frac{\int d\pi \int d\phi \exp \left( -\beta E[\phi ,\pi ]\right) \pi (x)}{{\cal Z}},\]

\[
\left\langle \pi (x)\pi (y)\right\rangle _{T}=\frac{\int d\pi \int d\phi \exp \left( -\beta E[\phi ,\pi ]\right) \pi (x)\pi (y)}{{\cal Z}},\]
which, after isolating the \( \phi  \) integral and canceling common terms
with \( {\cal Z} \), reduce to the following expressions:
\[
\left\langle \pi (x)\right\rangle _{T}=\frac{\int d\pi \exp \left( -\beta E_{0}[\pi ]\right) \pi (x)}{{\cal Z}_{0}},\: \: \: \: \left\langle \pi (x)\pi (y)\right\rangle _{T}=\frac{\int d\pi \exp \left( -\beta E_{0}[\pi ]\right) \pi (x)\pi (y)}{{\cal Z}_{0}}\]
with:
\[
E_{0}[\pi ]=\int _{0}^{L}dx\frac{1}{2}\pi ^{2}\: \: \: \mbox {and}\: \: \: {\cal Z}_{0}=\int d\pi \exp \left( -\beta E_{0}[\pi ]\right) \]
In order to compare these expectation values with the numerical ones (i.e.,
in the lattice), we evaluate the above expressions using the discrete version
with lattice spacing \( a \):
\[
\left\langle \pi (x)\right\rangle _{T}=\frac{1}{{\cal Z}_{0}}\int ...\int ^{+\infty }_{-\infty }\prod _{i}d\pi _{i}\exp \left[ -\beta a\sum _{j}\left( \frac{1}{2}\pi _{j}^{2}\right) \right] \pi _{x},\]

\[
\left\langle \pi (x)\pi (y)\right\rangle _{T}=\frac{1}{{\cal Z}_{0}}\int ...\int ^{+\infty }_{-\infty }\prod _{i}d\pi _{i}\exp \left[ -\beta a\sum _{j}\left( \frac{1}{2}\pi _{j}^{2}\right) \right] \pi _{x}\pi _{y}.\]
Let us begin with the expectation value \( \left\langle \pi (x)\right\rangle _{T} \),
which after some simplifications with \( {\cal Z}_{0} \) reduces to:
\begin{equation}
\label{expec_pi}
\left\langle \pi (x)\right\rangle _{T}=\frac{\int _{-\infty }^{+\infty }d\pi _{x}\exp \left( -\frac{\beta a}{2}\pi _{x}^{2}\right) \pi _{x}}{\int _{-\infty }^{+\infty }d\pi _{x}\exp \left( -\frac{\beta a}{2}\pi _{x}^{2}\right) }=0,
\end{equation}
where we have used the fact that we are integrating an odd function in a symmetric
interval. The two-point correlation function is obtained likewise:
\[
\left\langle \pi (x)\pi (y)\right\rangle _{T}=\frac{1}{{\cal Z}_{0}}\int ...\int ^{+\infty }_{-\infty }\prod _{i}d\pi _{i}\exp \left[ -\beta a\sum _{j}\left( \frac{1}{2}\pi _{j}^{2}\right) \right] \pi _{x}\pi _{y}=\]

\begin{equation}
\label{corr_pi}
=\frac{\int \int _{-\infty }^{+\infty }d\pi _{x}d\pi _{y}\exp \left[ -\frac{\beta a}{2}\left( \pi _{x}^{2}+\pi _{y}^{2}\right) \right] \pi _{x}\pi _{y}}{\int \int _{-\infty }^{+\infty }d\pi _{x}d\pi _{y}\exp \left[ -\frac{\beta a}{2}\left( \pi _{x}^{2}+\pi _{y}^{2}\right) \right] }=\frac{T}{a}\delta _{xy},
\end{equation}
where we took \( k_{B}=1 \) so that \( \beta =1/T \) and have used the following
general result for gaussian integrals:
\begin{equation}
\label{two_point_general}
\left\langle v_{n}v_{m}\right\rangle =\frac{1}{Z}\int ...\int ^{+\infty }_{-\infty }d^{N}\vec{v}\exp \left[ -\frac{1}{2}\sum _{i,j}A_{ij}v_{i}v_{j}\right] v_{n}v_{m}=\left( A^{-1}\right) _{nm},
\end{equation}
with \( A_{ij} \) a symmetric matrix and \( Z \) a normalization constant
(see, for instance, \cite{b_simons}). 

As in \cite{aarts2}, we shall study observables constructed from the following
two generating functions:
\begin{equation}
\label{corr_functions}
G^{(2)}(t)=\frac{1}{N}\sum _{x}\pi ^{2}(x,t),\: \: \: \: G^{(4)}(t)=\frac{1}{N}\sum _{x}\pi ^{4}(x,t)
\end{equation}

We now present a useful result for the correlation functions constructed from
these generating functions in \emph{equilibrium}:
\begin{equation}
\label{rel_g2_g4}
\left\langle G^{(4)}\right\rangle _{T}=\left\langle \pi ^{4}\right\rangle _{T}=3\left\langle \pi ^{2}\right\rangle _{T}^{2}=3\left\langle G^{(2)}\right\rangle _{T}^{2},
\end{equation}
which could be easily derived using the following relation for gaussian integrals
\cite{b_simons}:
\[
\left\langle v_{n}v_{m}v_{p}v_{q}\right\rangle =\left( A^{-1}\right) _{nm}\left( A^{-1}\right) _{pq}+\left( A^{-1}\right) _{np}\left( A^{-1}\right) _{mq}+\left( A^{-1}\right) _{nq}\left( A^{-1}\right) _{pm},\]
where \( A \) is the same matrix which appears in (\ref{two_point_general}).

We emphasize that these results were obtained in the canonical formalism, and
are therefore restricted to either a system coupled to a heat bath, or an isolated
system in the thermodynamical limit, where the canonical and the microcanonical
ensembles are equivalent. Otherwise these results should be taken only as an
approximation which becomes better for larger ensembles, i.e., close to the
\( N\rightarrow \infty  \) limit.

\section{The Lattice Formulation}

We have obtained the lattice version of the continuum \( \lambda \phi ^{4} \)
theory using second-order approximations for the derivatives in the equation
of motion (\ref{eq_motion}) (incidentally I also discuss these discretizations
with some details elsewhere \cite{adib}). We have then the \emph{leap-frog}
scheme:
\[
\frac{\phi _{i}^{n+1}-2\phi _{i}^{n}+\phi _{i}^{n-1}}{\Delta t^{2}}=\frac{\phi _{i+1}^{n}-2\phi _{i}^{n}+\phi _{i-1}^{n}}{\Delta x^{2}}-m^{2}\phi _{i}^{n}-\frac{1}{2}\lambda \left( \phi _{i}^{n}\right) ^{3}\Rightarrow \]

\begin{equation}
\label{leap-frog}
\phi _{i}^{n+1}=\rho \left( \phi _{i+1}^{n}+\phi _{i-1}^{n}\right) +2(1-\rho )\phi _{i}^{n}-\phi _{i}^{n-1}-\Delta t^{2}\left[ m^{2}\phi _{i}^{n}+\frac{1}{2}\lambda \left( \phi _{i}^{n}\right) ^{3}\right] ,
\end{equation}
with \( \rho =(\Delta t/\Delta x)^{2} \). The above equation could then be
solved iteratively for later times explicitly in terms of known values at times
\( n \).

We have solved equation (\ref{leap-frog}) with periodic boundary conditions,
together with the following (out-of-equilibrium) initial values for \( \phi  \)
and \( \pi  \) (again from \cite{aarts2}):
\begin{equation}
\label{init_values_cont}
\phi (x,0)=0,\: \: \: \: \pi (x,0)=\sum _{k}^{(n_{e})}A\cos (2\pi kx/L-\psi _{k}),
\end{equation}
where \( \psi _{k} \) is a random number in the range \( [0,2\pi ) \), and
the sum over \( k \) has \( n_{e} \) terms such that we only excite momenta
of the order of the mass. The above initial conditions could be implemented
using a first-order retarded difference scheme approximation for \( \pi =\partial _{t}\phi  \),
furnishing us with the following discrete variables:
\begin{equation}
\label{init_values_disc}
\phi ^{0}_{i}=0,\: \: \: \: \phi _{i}^{-1}=-\Delta t\sum _{k}^{(n_{e})}A\cos (2\pi kx/L-\psi _{k}).
\end{equation}

We have also tried a number of different ratios \( \Delta t/\Delta x \) in
order to guarantee the stability of the solutions since the \emph{standard}
Courant condition \emph{}\cite{adib} cannot be naively applied here (due to
the additional terms of first and third order). We have found stable solutions
for \( \rho =1/4 \) with \( m=0.001 \) and \( \lambda  \) ranging from \( \approx 0 \)
to \( 1000 \).

\section{Observables}

We simulate our microcanonical ensemble by generating \( N_{c} \) ``copies''
of the system with a given fixed energy \( E \), which depends only on the
initial amplitude \( A \), on the linear dimension \( L \) and on the number
of excited modes \( n_{e} \), as can be seen by equations (\ref{energy}) and
(\ref{init_values_cont}). 

In order to have a time-dependent expectation value for the observable \( O \)
in this ensemble, we take the average over the copies at each instant \( t \):
\[
\left\langle O\right\rangle _{N_{c}}(t)=\frac{1}{N_{c}}\sum _{n=1}^{N_{c}}O_{n}(t)\]

Since we are interested in the approach to equilibrium, we shall choose observables
that could be easily compared with the equilibrium ones. In our simulations
we have restricted ourselves to the function \( \mbox {dev}(\pi ) \) defined
in \cite{aarts2}, which is a measure of how much \( \pi  \) deviates from
being gaussian. Using the already shown relation (\ref{rel_g2_g4}) and the
following definition, we see that \( \mbox {dev}(\pi ) \) is a good measure
of ``how far'' we are from equilibrium:
\begin{equation}
\label{dev_pi}
\mbox {dev}(\pi )=\frac{\left\langle G^{(4)}\right\rangle _{N_{c}}(t)}{3\left\langle G^{(2)}\right\rangle _{N_{c}}^{2}(t)}-1,
\end{equation}
with \( \mbox {dev}(\pi )\rightarrow 0 \) as the thermalization is attained.

\section{The KAM Theorem}

A useful theorem in the study of the ergodicity of finite systems is the \emph{KAM
theorem} (see, for instance, \cite{parisi, casetti}). We will follow these
two references and stick to the qualitative view of the theorem, which roughly
says that if the Hamiltonian of our finite system has an integrable term \( {\cal H}_{0} \)
(i.e., it has some integrals of motion) plus a smooth perturbation \( g{\cal H} \),

\[
H={\cal H}_{0}+g{\cal H},\]
then for sufficiently small \( g \) the system will \emph{not} be ergodic.
This might suggest that there is a certain critical value of \( g \) which
separates sharply the ergodic and non-ergodic regimes, and indeed it was found
in some numerical simulations of the FPU problem \cite{parisi}. However, as
we will show later, our system presents a smooth divergent behavior of the relaxation
time as \( \lambda \rightarrow 0 \), thus excluding the existence of the above
critical value.

The gaussian model, which is obtained by taking \( \lambda =0 \) in our theory,
has a closed-form solution with constants of motion and is therefore restricted
to sweep a very limited area in the phase space. Thus, the ergodicity condition
\begin{equation}
\label{ergodicity}
\lim _{t\rightarrow \infty }\int _{0}^{t}dt'g[C(t')]\rightarrow \frac{1}{\Omega }\int d\mu (C)\delta \left( E-H(C)\right) g(C)
\end{equation}
with \( d\mu (C) \) the measure of the configuration space and \( \Omega  \)
a normalization constant, is not satisfied, so this system is not a ``good''
statistical ensemble.

However, in the non-linear \( \lambda \neq 0 \) case (i.e., with a term of
order \( \phi ^{3} \) in addition to the linear term in the equation of motion),
we might expect ergodicity for some values of \( E \) and \( \lambda  \),
and in fact this system has already been studied also by Parisi in the context
of a chain of anharmonically coupled oscillators (the well-known Fermi-Pasta-Ulam
problem \cite{parisi2}). It is found in this latter paper that the system doesn't
present the critical \( g \) said above; on the contrary, it \emph{always}
relax towards equilibrium, although with divergent time scales as \( g\rightarrow 0 \).
Here we shall also analyze these issues, albeit with a different approach.

\section{Simulation Results and Discussions }

\begin{figure}
{\centering \resizebox*{0.5\textwidth}{!}{\rotatebox{-90}{\includegraphics{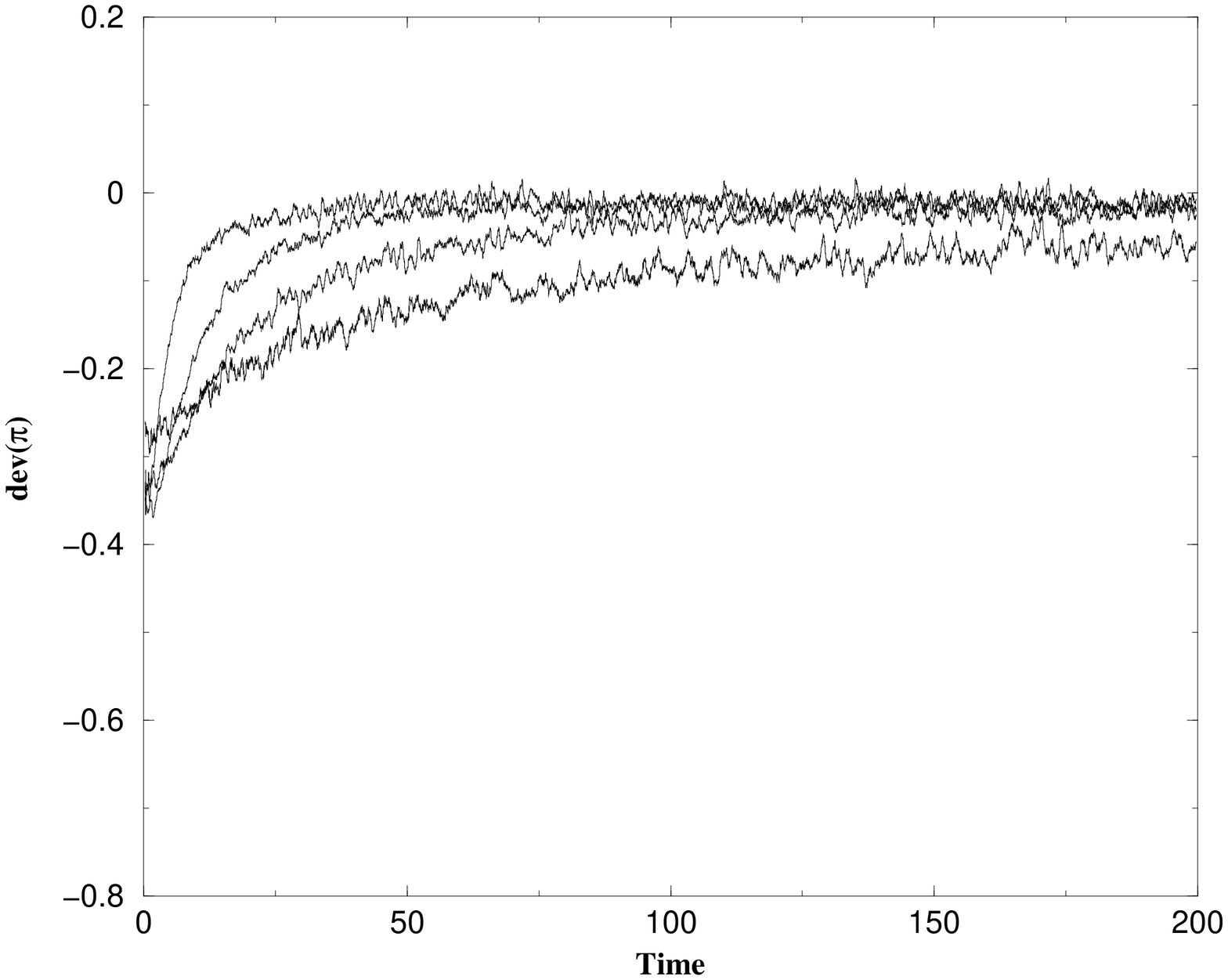}}} \par}

\caption{\label{fig: figura1}Time-dependence of dev(\protect\( \pi \protect \)) for
different values of \protect\( \lambda \protect \). From top to bottom: \protect\( \lambda =1000,100,10,1\protect \).
These results were run for an ensemble of \protect\( N_{c}=10\protect \), all
of which with \protect\( N=200\protect \) lattice points. In order to make
the graph clearer, we have reduced the fluctuations by taking local time averages
of length \protect\( 50\protect \) time-steps.}
\end{figure}
\begin{figure}
{\centering \resizebox*{0.5\textwidth}{!}{\includegraphics{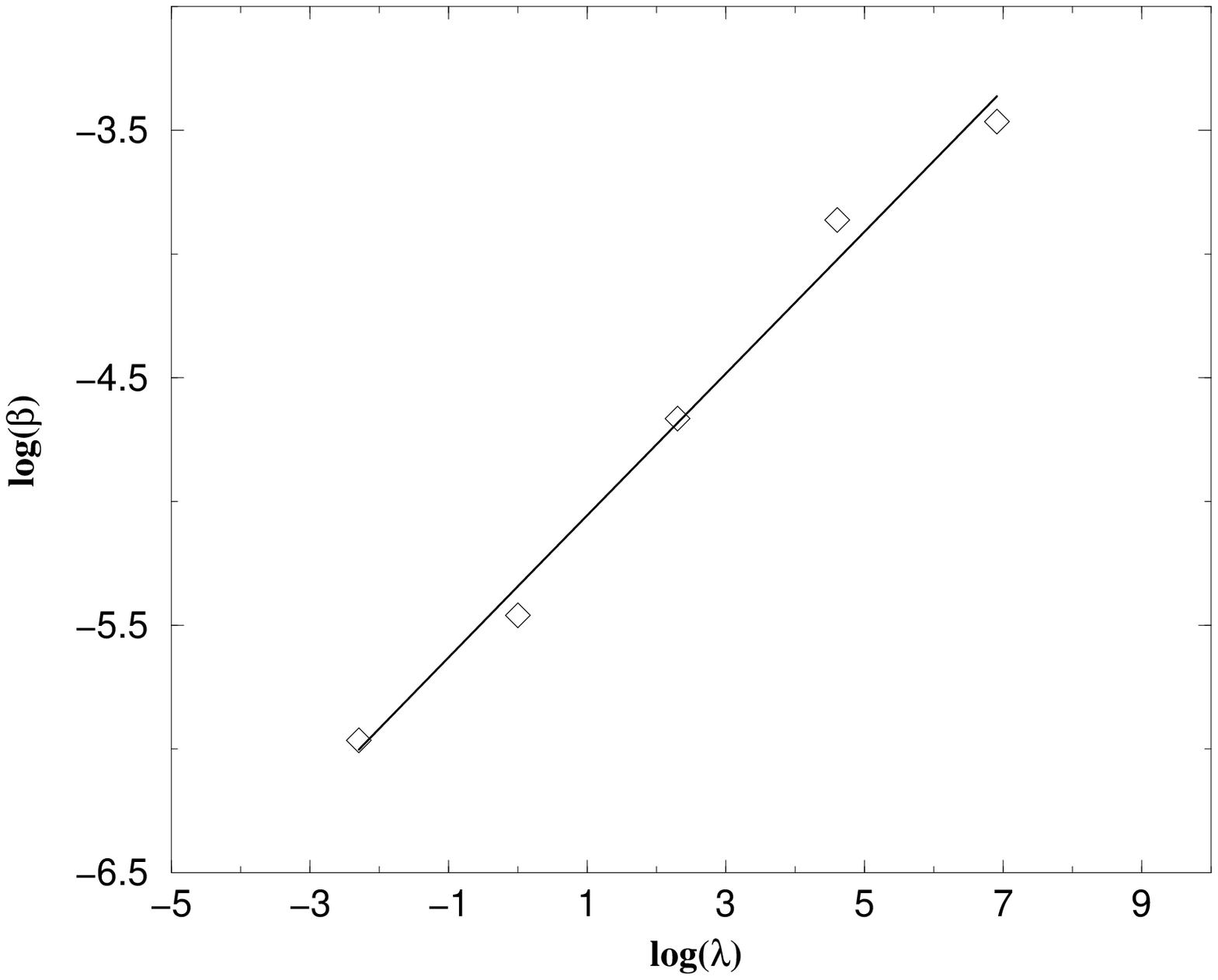}} \par}

\caption{\label{fig: figura2}Behavior of the slope \protect\( \beta \protect \) (see
Figure \ref{fig: figura3}) for different values of \protect\( \lambda \protect \).
We see here both the calculated values (circles) and the linear fitting. Since
\protect\( \beta \protect \) measures the rapidity with which the thermalization
is attained (at least initially), we see that the relaxation time diverges as
\protect\( \lambda \rightarrow 0\protect \).}
\end{figure}

\begin{figure}
{\centering \resizebox*{0.5\textwidth}{!}{\includegraphics{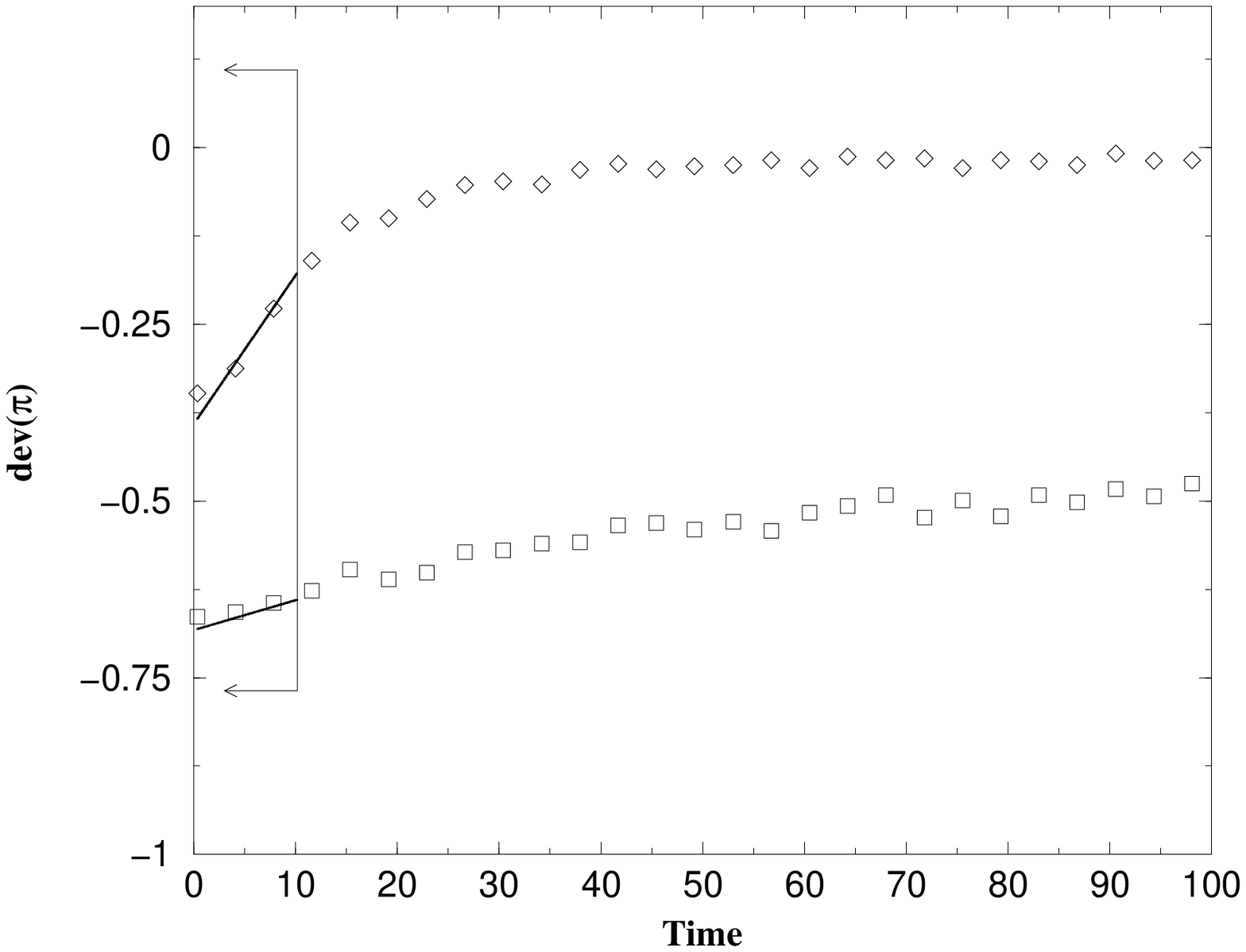}} \par}

\caption{\label{fig: figura3}How to obtain \protect\( \beta \protect \) from Figure
\ref{fig: figura1}. The symbols represent the numerical results (with local
time averages), whereas the lines are linear fittings. The ``rapidity'' \protect\( \beta \protect \)
is defined as the slope of the linear fitting from \protect\( t=0\protect \)
to \protect\( 10\protect \). The upper curve is for \protect\( \lambda =100\protect \)
and the lower one is for \protect\( \lambda =1\protect \) (the lower curve
has been shifted down by \protect\( 0.4\protect \) in order to make the graph
clearer).}
\end{figure}

 We have run the simulations on the lattice described previously for different
values of \( \lambda  \) and \( N \) (we have fixed both the lattice length
\( L=1 \) and the mass term \( m=0.001 \), since these values together with
\( \lambda  \) ranging from \( 1 \) to \( 1000 \) allowed the system to behave
in very different ways, thermalizing within \( t=100 \) for \( \lambda \approx 1000 \)
and being almost locked in the initial phase for \( \lambda \approx 1 \)).
The results for different values of the coupling constant are shown in Figures
\ref{fig: figura1} and \ref{fig: figura2} (the slope \( \beta  \) used in
Figure \ref{fig: figura2} is explained in Figure \ref{fig: figura3}). We notice
that the relaxation times depend strongly on the coupling constant, being the
faster times for greater \( \lambda  \).

These results are in good agreement with \cite{parisi2}, which found no critical
value for the coupling constant (albeit in the already mentioned context of
the FPU problem) and also discuss the apparent existence of this critical value
when one interprets naively the relaxation times in similar plots. We see from
Figure \ref{fig: figura2} that the ``rapidity'' \( \beta  \) vanishes in
a power-law as \( \lambda \rightarrow 0 \). Since in all of our numerical studies
the function dev(\( \pi  \)) presented an almost monotonic increasing behavior,
we found that \( \beta  \), although not a precise measure of the relaxation
time, has proven to be a very elucidative tool in estimating the time-scales
in our restricted spectra of parameters.

\begin{figure}
{\centering \resizebox*{0.5\textwidth}{!}{\includegraphics{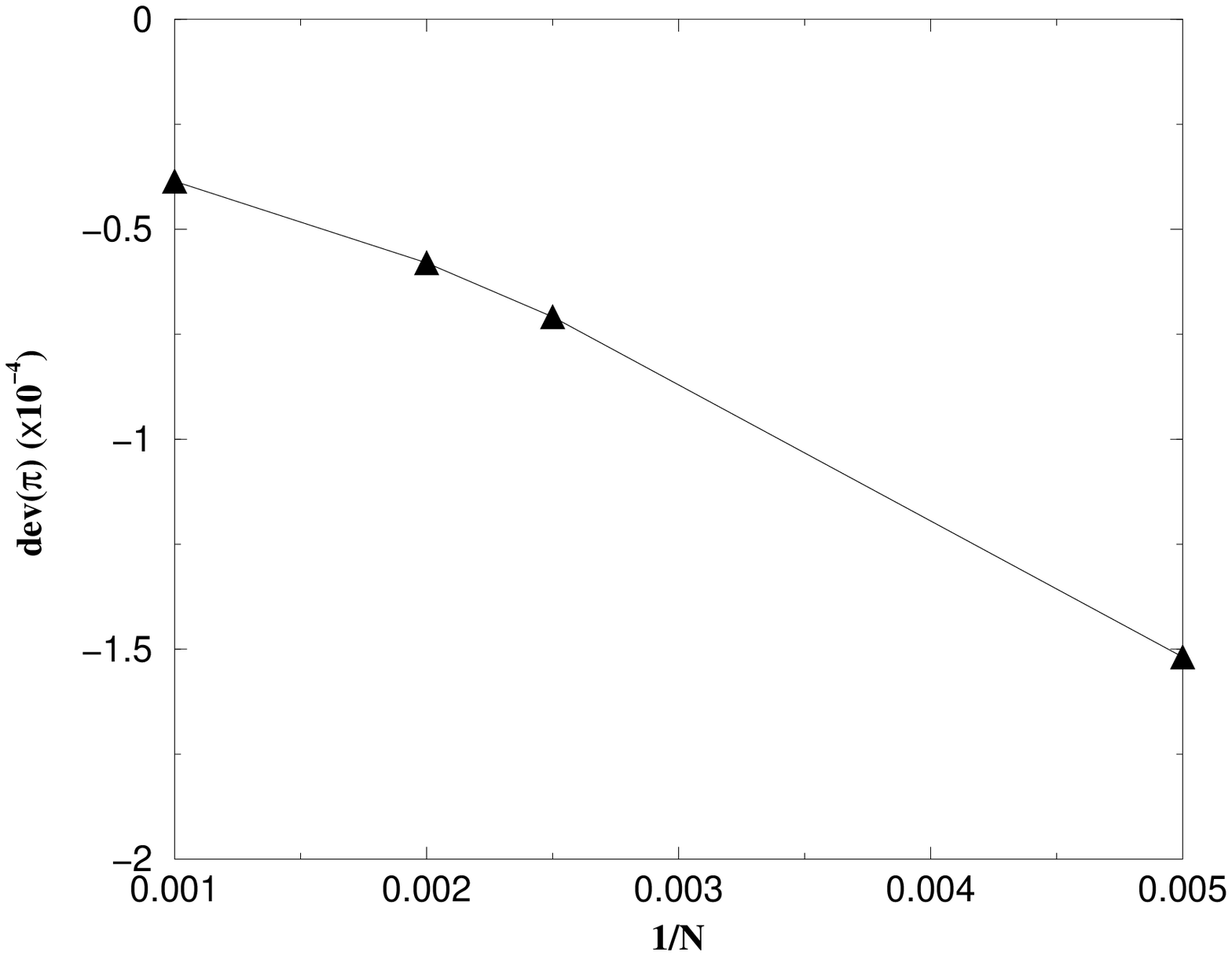}} \par}

\caption{\label{fig: figura4}dev(\protect\( \pi \protect \)) in the quasi-stationary
regime (\protect\( 180<t<200\protect \), \protect\( \lambda =1000\protect \))
for different values of \protect\( N\protect \). The triangles show the numerical
results whereas the lines are straight-line interpolations.}
\end{figure}

We have also studied our system with regard to its dependence on the number
of sites \( N \). We remember that dev(\( \pi  \)) was defined using canonical
expectation values, whereas our system is a completely closed (microcanonical)
ensemble. We justify this approach by noting that in the thermodynamic limit
(i.e., \( N\rightarrow \infty  \)) the statistical ensembles should be equivalent
(except, perhaps, for very rare systems which we won't touch here). We illustrate
this behavior in Figure \ref{fig: figura4}, where we can see that our microcanonical
dev(\( \pi  \)) tends to the canonical zero value as \( N\rightarrow \infty  \).
The values used in this figure were obtained by doing an average in the ``quasi-stationary''
regime (following Aarts), where the system doesn't show any tendency of changing
the expectation value of dev(\( \pi  \)).

\section{Conclusions}

We have studied numerically the approach to equilibrium of the \( \lambda \phi ^{4} \)
theory, showing the strong dependence of the relaxation times on the coupling
constant \( \lambda  \). In order to analyze how far the theory is from equilibrium,
we have adopted an observable which could be defined locally in time during
the simulation and also easily calculated in equilibrium using the standard
canonical probability measure, depending essentially on the conjugate momentum
\( \pi  \).

We have also defined a quantity \( \beta  \) (which we called ``rapidity'')
which measures how fast our theory is approaching to equilibrium. This quantity,
as defined here, does \emph{not} measure the overall behavior of the thermalization,
but furnishes us with a good illustration of the divergent behavior of the relaxation
time as the coupling constant vanishes. This result is not, at first sight,
to be expected when one looks at the KAM theorem, since it might suggest a critical
behavior of the relaxation times with \( \lambda  \).

We finish our study by showing how the system behaves for large values of the
number of sites \( N \), finding the vanishing of the quasi-stationary expectation
value of dev(\( \pi  \)) as \( N\rightarrow \infty  \). This latter result
justifies the use of the canonical probability measure to calculate the equilibrium
expectation values.

\section{Acknowledgements}

I would like to thank Prof. Carlos Alberto S. Almeida for his generosity in
accepting me for this undergraduate research program even though it is in a
different branch from what he has been doing, Prof. Murilo Pereira de Almeida
for his kind discussions, Dr. Gert Aarts for the prompt help, and Prof. Marcelo
Gleiser for his stimulating letters, without which I wouldn't have pursued my
studies on this subject.

\end{document}